\documentstyle[aps,prl,tighten,12pt,psfig]{revtex}

\newtheorem{theorem}{Theorem}
\newtheorem{lemma}{Lemma}

\begin{document}

\title{Taylor expansion for an operator function}
\author{Ioan Sturzu \\
%EndAName
Center for Computational Nanoscience,\\
Ball State University, Muncie, IN 47306} \maketitle

\begin{abstract}
A simple version for the extension of the Taylor theorem to the
operator functions was found. The expansion was done with respect to
a value given by a diagonal matrix for the non-commutative case, and
the coefficients are given both by recurrence relations and Cauchy
integrals.
\end{abstract}
\vspace*{1cm}

 In Quantum Physics, one has to use functions of the operators
defined in some Hilbert spaces to describe specific quantities, as
evolution operator, state operator. Usually, these are known for
some more simple operators, while changing some parameters, the
problems become analytically intractable, usually because the
involved operators cannot be easy diagonalizable anymore. However,
approximate solutions exist when the new operators differ (in some
appropriate norm) from the analytically-tractable ones. Here we
present two independent mathematical proofs  of a useful result,
which straightforwardly generalizes the classical result of Taylor
expansion from the real analysis to  the operator algebra domain.

\vspace*{0.5cm}
 A function of an operator defined on a Hilbert space
can be written using a Cauchy-type formula [1]:

\begin{equation}\label{funcop}
    f(A)=\frac 1{2 \pi i}\oint\limits_\gamma \frac{f(z)}{z-A}
\end{equation}
where $\gamma$ is the closed rectifiable Jordan boundary of an open
domain containing the spectrum of the operator $A$.

 The following lemma applies:

\begin{lemma}
\label{lemma}If $\lambda $ is a diagonal matrix, $\left[ \lambda \right]
_{ip}=\lambda _i\delta _{ip}$, than:

\begin{equation}
\left( {\lambda +\tau }\right) ^p={\lambda }^p(1+\sum_{q=0}^{p-1} \epsilon
_q+...\sum_{q=0}^{p-1}\sum_{q_1=q+1}^{p-1}...\sum_{q_{k-1}=q_{k-2}+1}^{p-1}\epsilon
_{q_{k-1}}...\epsilon _{q_1}\epsilon _q+...)
\end{equation}
where:
\begin{equation}
\left[ \epsilon _q\right] _{ip}=\left( \frac{{\lambda _p}}{{\lambda _i}} \right) ^q\frac
1{{\lambda _i}}\tau _{ip}
\end{equation}
\end{lemma}

\begin{description}
\item[Proof]  One has:
\[
\left[ {\lambda }\epsilon _{q_1}...\epsilon _{q_r}{\lambda }\right] _{ip}{ =\lambda
}_i{\sum\limits_{m_1,m_2,...m_{r-1}}}\left[ \epsilon _{q_1}\right] _{im_1}\left[ \epsilon
_{q_2}\right] _{m_1m_2}...\left[ \epsilon _{q_r}\right] _{m_{r-1}p}{\lambda }_p=
\]
\[
={\lambda }_i{\sum\limits_{m_1,m_2,...m_{r-1}}}\left[ \epsilon _0\right] _{im_1}\left[
\epsilon _0\right] _{m_1m_2}...\left[ \epsilon _0\right] _{m_{r-1}p}\left( \frac{{\lambda
_{m_1}}}{{\lambda _i}}\right) ^{q_1}\left( \frac{{\lambda _{m_2}}}{{\lambda _{m_1}}}\right)
^{q_2}...\left( \frac{{ \lambda _p}}{{\lambda _{m_r-1}}}\right) ^{q_r}{\lambda }_p=
\]
\[
={\lambda }_i^2{\sum\limits_{m_1,m_2,...m_{r-1}}}\left[ \epsilon _0\right]
_{im_1}\left[ \epsilon _0\right] _{m_1m_2}...\left[ \epsilon _0\right]
_{m_{r-1}p}\left( \frac{{\lambda _{m_1}}}{{\lambda _i}}\right)
^{q_1+1}\left( \frac{{\lambda _{m_2}}}{{\lambda _{m_1}}}\right)
^{q_2+1}...\left( \frac{{\lambda _p}}{{\lambda _{m_r-1}}}\right) ^{q_r+1}{=}
\]
\begin{equation}
={\lambda }_i^2{\sum\limits_{m_1,m_2,...m_{r-1}}}\left[ \epsilon
_{q_1+1}\right] _{im_1}\left[ \epsilon _{q_2+1}\right] _{m_1m_2}...\left[
\epsilon _{q_r+1}\right] _{m_{r-1}p}=\left[ {\lambda }^2\epsilon
_{q_1+1}...\epsilon _{q_r+1}\right] _{ip}
\end{equation}
The proof is done by mathematical induction:
\[
\left( {\lambda +\tau }\right) ^1={\lambda }(1+\epsilon _0)
\]
\[
\left( {\lambda +\tau }\right) ^{p+1}={\lambda }^p(1+\sum_{q=0}^{p-1}\epsilon
_q+...\sum_{q=0}^{p-1}\sum_{q_1=q+1}^{p-1}...\sum_{q_{k-1}=q_{k-2}+1}^{p-1}\epsilon
_{q_{k-1}}...\epsilon _{q_1}\epsilon _q+...){\lambda }(1+\epsilon _0)=
\]
\[
={\lambda }^{p+1}(1+\sum_{q=1}^p\epsilon _q+...\sum_{q=1}^p\sum_{q_1=q+1}^p...\sum
_{q_{k-1}=q_{k-2}+1}^p\epsilon _{q_{k-1}}...\epsilon _{q_1}\epsilon _q+...)(1+\epsilon _0)=
\]
\[
={\lambda }^{p+1}(1+\sum_{q=0}^p\epsilon _q+...\sum_{q=0}^p\sum_{q_1=q+1}^p...\sum
_{q_{k-1}=q_{k-2}+1}^p\epsilon _{q_{k-1}}...\epsilon _{q_1}\epsilon _q+...)
\]
\end{description}

\begin{theorem}
Let $f$ be a real function, consistent with the conditions of the usual Taylor expansion
theorem. For the matrices ${\lambda }${\ and }${\tau }$ as in Lemma \ref{lemma}, one can
construct the following Taylor expansion with respect to the value given by the
 diagonal matrix $\lambda$:

\[
\left[ {f(\lambda +\tau )}\right] _{ip}=f(\lambda _i)\delta _{ip}+\sum\limits_{n\ge
1}{\sum\limits_{m_1,m_2,...m_{n-1}}{ A_{i,m_1,m_2,...m_{n-1},p}^{(n,\lambda )}}}\tau
_{im_1}\tau _{m_1m_2}...\tau _{m_{n-1}p}
\]
where:
\begin{equation}
A_{ip}^{(1,\lambda )}=\left\{
\begin{array}{l}
\frac{{{f(\lambda }_i{)}-{f(\lambda }_p{)}}}{{\lambda _i-\lambda _p}},\text{
if }{\lambda _i\neq \lambda _p} \\
\\
{f}^{\prime }{{(\lambda }_i{),}}\text{ else}
\end{array}
\right. \label{recc0}
\end{equation}

\[
A_{i,m,p}^{(2,\lambda )}=\left\{
\begin{array}{l}
\frac{A_{ip}^{(1,\lambda )}{-}A_{mp}^{(1,\lambda )}}{{\lambda _i-\lambda _m}}
,\text{ if }{\lambda _i\neq \lambda _m} \\
\\
\text{else limit, i.e. } \left\{\begin{array}{l} \frac{{{f(\lambda }_i{)}-{f(\lambda
}_p{)}}}{\left( {\lambda _i-\lambda _p} \right) ^2}-\frac{{f}^{\prime }{{(\lambda
}_p{)}}}{{\lambda _i-\lambda _p}},
\text{ if }{\lambda _i\neq \lambda _p} \\
\\
\frac 12{f}^{\prime \prime }{{(\lambda }_i{),}}\text{ else}
\end{array}
\right.
\end{array}
\right. \label{recc1} \]

\begin{equation}
A_{i,m_1,m_2,...m_{n-1},m_n,p}^{(n+1,\lambda )}= \left\{\begin{array}{l}
\frac{{A_{i,m_2,...m_{n-1},p}^{(n,\lambda )}-A_{m_1,m_2,...m_{n-1},p}^{(n,\lambda
)}}}{{\lambda _i-\lambda _{m_1}}},
\text{ if }{\lambda _i\neq \lambda _p} \\
\\
\text{else limit}
\end{array}
\right.
\label{recc}
\end{equation}
\end{theorem}

\begin{description}
\item[Proof] Using (\ref{funcop}) one can can
write a Taylor expansion  with respect to the null operator [1]:
\begin{eqnarray}
{f(\lambda +\tau )} &=&{}\sum_{n\geq 0}\frac 1{n!}{f}^{\prime (n)}{(0)} \left( {\lambda
+\tau }\right) ^n=\sum_{n\geq 0}\frac 1{n!}{f}^{\prime (n)}{
(0)\lambda }^n  \nonumber \times \\
&& \times
(1+...\sum_{q=0}^{p-1}\sum_{q_1=q+1}^{p-1}...\sum_{q_{k-1}=q_{k-2}+1}^{p-1}\epsilon
_{q_{k-1}}...\epsilon _{q_1}\epsilon _q+...)
\end{eqnarray}
the $(i,p)$ matrix element of the $k$ term from the parenthesis can be written as:
\begin{equation}
{\sum\limits_{m_1,m_2,...m_{k-1}}B_{i,m_2,...m_{k-1},p}^{(k,\lambda ,n)}} \tau _{im_1}\tau
_{m_1m_2}...\tau _{m_{n-1}p}
\end{equation}
where:
\begin{eqnarray*}
{B_{i,m_1,m_2,...m_{k-1},p}^{(k,\lambda ,n)}} &=&{}\sum_{q=0}^{n-1}
\sum_{q_1=q+1}^{n-1}...\sum_{q_{k-1}=q_{k-2}+1}^{n-1}\left( \frac{{\lambda
_{m_1}}}{{\lambda _i}}\right) ^{q_{k-1}}\left( \frac{{\lambda
_{m_2}}}{{\lambda _{m_1}}}\right) ^{q_{k-2}}... \\
&&...\left( \frac{{\lambda _{m_k-1}}}{{\lambda _{m_k-2}}}\right) ^{q_1}\left(
\frac{{\lambda _p}}{{\lambda _{m_k-1}}}\right) ^q\frac 1{{ \lambda _i\lambda
_{m_1}...\lambda _{m_k-1}}}
\end{eqnarray*}

 In the corresponding expression for ${B_{i,m_1,m_2,...m_k,p}^{(k+1, \lambda ,n)}}$,
if ${\lambda _i\neq \lambda _{m_1}}$ one has:
\begin{equation}
\sum_{q_k=q_{k-1}+1}^{n-1}\left( \frac{{\lambda _{m_1}}}{{\lambda _i} }\right)
^{q_{k-1}}=\left( \frac{{\lambda _{m_1}}}{{\lambda _i}}\right) ^{q_{k-1}+1}\frac{1-\left(
\frac{{\lambda _{m_1}}}{{\lambda _i}}\right) ^{n-1-q_{k-1}}}{1-\frac{{\lambda
_{m_1}}}{{\lambda _i}}}  \label{sum}
\end{equation}
so, one obtains after some algebra:
\begin{equation}
{B_{i,m_1,m_2,...m_k,p}^{(k+1,\lambda ,n)}=}\frac{{B_{i,m_2,...m_k,p}^{(k, \lambda
,n)}-\left( \frac{{\lambda _{m_1}}}{{\lambda _i}}\right) ^nB_{m_1,m_2,...m_k,p}^{(k,\lambda
,n)}}}{{\lambda _i}-{\lambda _{m_1}}} \label{recc1}
\end{equation}
Identifying the coefficients:
\begin{equation}
A_{i,m_1,m_2,...m_{k-1},p}^{(k,\lambda )}=\sum_{n\geq 0}\frac 1{n!}{f} ^{\prime
(n)}{(0)\lambda }_i^n{B_{i,m_1,m_2,...m_{k-1},p}^{(k,\lambda ,n)}}
\end{equation}
and using (\ref{recc1}), one obtains the recurrence relation (\ref{recc}).
If ${\lambda _i=\lambda _{m_1}}$, (\ref{sum}) is equal to $(n-1-q_{k-1})$;
finally the term $n$ will go to a derivative (which is finite, because $f$
is Taylor), while the contribution of the remainder is also finite. So, for
the higher orders, the limiting argument is valid, if for the lower orders
this is true. However, for orders $1$ and $2$ this was verified by direct
calculation, so the theorem is true, by the argument of mathematical
induction.
\end{description}

The same result can be obtained starting with a Dyson expansion of
the integrand from (\ref{funcop}).

\begin{equation}
\frac 1{z-\left( {\lambda +\tau }\right) }=\frac 1{z-{\lambda }%
}+\sum\limits_{n\ge 1}(\frac 1{z-{\lambda }}{\tau )}^n\frac
1{z-{\lambda }}
\end{equation}
Using (\ref{funcop}) one has:
\begin{eqnarray}
\left[ {f(\lambda +\tau )}\right] _{ip} &=&f(\lambda _i)\delta
_{ip}+\sum\limits_{n\ge 1}{\sum\limits_{m_1,m_2,...m_{n-1}}}%
\oint\limits_\gamma \frac{{f(z)dz}}{\left( z-{\lambda }_i\right) \left( z-{%
\lambda }_{m_1}\right) ...\left( z-{\lambda }_{m_{n-1}}\right)
}\times\\
&&\times \tau _{im_1}\tau _{m_1m_2}...\tau _{m_{n-1}p}
\end{eqnarray}

Identifying:
\begin{equation}
{{A_{i,m_1,m_2,...m_{n-1},p}^{(n,\lambda )}=}}\oint\limits_\gamma \frac{{%
f(z)dz}}{\left( z-{\lambda }_i\right) \left( z-{\lambda
}_{m_1}\right) ...\left( z-{\lambda }_{m_{n-1}}\right) }
\end{equation}
after decomposing the path $\gamma$ in closed paths surrounding only
one of the values $\lambda$ one can obtain very easy the recurrence
relations (\ref{recc0}) and (\ref{recc1}).

\vspace*{0.5cm} \noindent References
\baselineskip=0.2cm

\begin{enumerate}
\item  N. Dunford and J. Schwartz, Linear Operators (Interscience Publ., 1957).

\end{enumerate}

\vspace*{0.5cm} This work has been supported by The Indiana 21st
Century Research and Technology Fund.
\end{document}